\def\half{\frac{1}{2}}
\def\nll{\smallskip\hfil\hfil\linebreak\noindent} 
\def\hi#1#2{$#1$\kern -2pt-#2} 
\def\mb#1{\mbox{\boldmath{$#1$}}}
\def\dbox#1{\hbox{\vrule 
\vbox{\hrule \vskip #1\hbox{\hskip #1\vbox{\hsize=#1}\hskip #1}\vskip #1\hrule}\vrule}}
\def\qed{\begin{flushright}~{\dbox{0.05true in}}\end{flushright}} 

\def\refeq#1{Eq.\,(\ref{#1})}
\def\extr{{\rm extr}}
\def\csch{{\rm csch}}
\def\sgn{{\rm sgn}}
\def\sech{{\rm sech}}
\documentclass[amsmath,amssymb,superscriptaddress,showkeys,showpacs,nofootinbib,10pt]{revtex4}
\usepackage{amsmath}
\usepackage{bm}
\usepackage{graphicx}
\usepackage{rotating}
\usepackage{mathrsfs} 
\newtheorem{theorem}{Theorem}
\newtheorem{example}{Example}
\newtheorem{proof}{Proof}
\usepackage{collectbox}
\makeatletter
\newcommand{\mybox}{%
    \collectbox{%
        \setlength{\fboxsep}{3pt}%
        \fbox{\BOXCONTENT}%
    }%
}

\begin{document}

\begin{flushright}  CUQM~159 ~~~~~~~\end{flushright}
\def\ttitle{Potential envelope theory and the local energy theorem}
\title{\ttitle}
\author{Ryan~Gibara and Richard L.~Hall}
\email{ryan.gibara@concordia.ca}
\email{richard.hall@concordia.ca}
\affiliation{Department of Mathematics and Statistics, Concordia University,
1455 de Maisonneuve Boulevard West, Montr\'eal,
Qu\'ebec, Canada H3G 1M8}

\begin{abstract}
We consider a one--particle bound quantum mechanical system governed by a Schr\"odinger operator $\mathscr{H} = -\Delta + v\,f(r)$, where $f(r)$ is an attractive central potential, and $v>0$ is a coupling  parameter.  If $\phi \in \mathcal{D}(\mathscr{H})$ is a `trial function', the local energy theorem tells us that the discrete energies of $\mathscr{H}$ are bounded by the extreme values of $(\mathscr{H}\phi)/\phi,$ as a function of $r$.  We suppose that $f(r)$ is a smooth transformation of the form $f = g(h)$, where $g$ is monotone increasing with definite convexity and $h(r)$ is a potential for which the eigenvalues $H_n(u)$ of the operator $\mathcal{H}=-\Delta + u\, h(r)$, for appropriate $u >0$, are known.  It is shown that the eigenfunctions of $\mathcal{H}$ provide local-energy trial functions $\phi$ which necessarily lead to finite eigenvalue approximations that are either lower or upper bounds.  This is used to extend the local energy theorem to the case of upper bounds for the excited-state energies when the trial function is chosen to be an eigenfunction of such an operator $\mathcal{H}$.  Moreover, we prove that the local-energy approximations obtained are identical to `envelope bounds', which can be obtained directly from the spectral data $H_n(u)$ without explicit reference to the trial wave functions.
\end{abstract}

\keywords{Envelope theory, local energy theorem, kinetic potentials} 

\pacs{03.65.Ge.}

{\let\newpage\relax\maketitle}
\section{Introduction}
We study the relationship between two methods of approximating the discrete spectrum of a Schr\"odinger operator $\mathscr{H} = -\Delta + V$.  We assume that $\mathscr{H}$ is bounded below and essentially self  adjoint on a suitable domain $\mathcal{D}(\mathscr{H}) \subset L^2(\mathbb{R}^d)$, and that it supports a discrete spectrum.  Furthermore, we shall consider attractive `spherically-symmetric' potential functions of the form $V(\mb{r}) = v f(r),$ where $v >0$ is a positive coupling parameter and $f(r)$ describes the shape of the potential along a radial line, where $r = ||\mb{r}||,\,\mb{r} \in\mathbb{R}^d$.  For simplicity of presentation we shall discuss many general aspects of the problem with reference to the one--dimensional case $d = 1$ for which the potential shape $f(x)$ is an even function of $x \in \mathbb{R}$.  An eigenvalue is written $E_n = F_n(v)$ and the corresponding wave function $\psi_{n}(x)$ has $n$ nodes, $n = 0,1,2,\dots.$  We study an example in detail in  $d>1$ dimensions in section 6.
\medskip 
 
In the present context the local energy theorem uses a `trial function' $\phi$ in a sense different from a variational analysis.  The following succinct statement of the theorem may be found in volume 3 of {\it A Course on Mathematical Physics} by Walter Thirring \cite{thirring}: 
\begin{equation}\label{let1}
 \min_{x > 0}\left(\frac{\mathscr{H}\phi(x)}{\phi(x)}\right) \, \leq \, E \, \leq \, \max_{x > 0}\left(\frac{\mathscr{H}\phi(x)}{\phi(x)}\right),
\end{equation}
where, for the ground state, $\phi(x)$ is a nodefree trial function.  As with variational estimates, the energy bounds can often be improved by exploring their dependence on parameters built in to the trial function $\phi.$  However, there is an important practical matter to face, namely that it seems very difficult to know ahead of time when the minimum is $-\infty$ or the maximum is $+\infty$, which extrema clearly add nothing new.  A pragmatic aspect of the present paper is that it provides a way of choosing $\phi$ so that it is guaranteed to yield either a finite lower bound or a finite upper bound.  However, interest in the solution of this problem may transcend its utility for selecting local-energy trial functions since it relates the local energy theorem to an established geometrical analysis called `potential envelope theory' that we shall show yields precisely the same energy bounds as \refeq{let1}.  Moreover, we can obtain both lower as well as upper bounds for every discrete eigenvalue: for the excited states, the extrema are found by using $\inf$ and $\sup$ while omitting the zeros of a suitably chosen $\phi(x)$.
\medskip 

We shall discuss the mathematical basis for the methods in the sections which follow, but it will help to fix ideas if we explain here what is the source of the trial wave functions $\phi$ that do have the claimed properties.  It is perhaps interesting that, after we have established the connection between the theories via $\phi$, we can go on to find the energy bounds themselves directly without using $\phi$ at all.  We consider all the discrete eigenvalues as functions of the coupling parameter $v >0,$ thus the operator we are studying and its unknown discrete eigenvalues $F_n(v)$ are given by 
\begin{equation}\label{Fn}
\mathscr{H} = -\Delta + v f(r) \, \longrightarrow \, \{E_n = F_n (v)\}_{n = 0,1,2,\dots}, \,
 v > 0.
\end{equation}
Meanwhile we suppose that the following operator with attractive potential shape $h(r)$ has known eigenvalues $H_n(u)$
\begin{equation}\label{base}
\mathcal{H} = -\Delta + u\,h(r) \, \longrightarrow \, \{\mathcal{H}\phi_n = H_n (u)\phi_n\}_{n=0,1,2\dots}, \, u >0.
\end{equation}
The assumed link between the problems is that the potential shape $f(r)$ has the representation $f(r) = g(h(r))$, where $g$ is an increasing function of $h(r)$ with definite convexity, thus
\begin{equation}\label{connection}
 f(r ) = g(h(r)), \, g'(h) > 0, \, g''(h) \ne   0.
\end{equation} 
The envelope prescription for the local-energy trial function $\phi_n(r)$ is that it is an eigenfunction of $-\Delta + u h(r)$ with eigenvalue $H_n(u).$  We claim that if this $\phi_n$ with suitable $u$ is inserted into \refeq{let1}, then it yields a lower bound to $E_n$ if $f = g(h)$ is convex ($g''> 0$) and an upper bound to $E_n$ if $f = g(h)$ is concave ($g''<0$).  This is the case for each choice of the coupling $v$ for which the eigenvalues exist.  For example, it is a well-known  feature of the problem that a discrete eigenvalue may only exist for an excited state when the coupling $v$ is sufficiently large.  
\medskip
 
Potential envelope theory and the local energy theorem are discussed in more detail in sections 2 and 3.  The coincidence of eigenvalues induced by the above prescription for $\phi$ and the derivation of the bounds directly from the input spectral data $H_n(u)$ are presented in sections 4 and 5.  A simple example with an oscillator envelope basis $h(x) = x^2$ is mentioned in each section, as a connecting thread.  The problem for central potentials in $d > 1$ dimensions is formulated in section 6 and an example with both upper and lower spectral bounds is presented in detail.

\section{Potential envelope theory}
The `method of potential envelopes' was introduced in 1980 by Hall \cite{hall1} who subsequently developed the idea into a spectral approximation and inversion theory \cite{hall3,hall4,hall5,hall6,hall7}. Some of the results were re-discovered (starting 28 years later)  by Buisseret,  Semay, and  Silvestre-Brac~\cite{buis} who called their approach `the auxiliary field method'.  For our present purpose, we make use of the assumed connection \refeq{connection} between the two Schr\"odinger operators $\mathscr{H}$ and $\mathcal{H}$ to obtain an approximation of $F_n(v)$ in terms of $H_n(u)$.  For definiteness we first take $g$ to be convex (i.e. $g''(h)>0$). Consider the family of all tangents to $g(h)$.  For each point of contact $x=t$, the tangential potential is of the form $a(t)+b(t)h(x)$ where 
\begin{displaymath}
\left\{\begin{array}{ll}a(t)=&g(h(t))-g'(h(t))h(t)\\b(t)=&g'(h(t)).\end{array}\right\}
\end{displaymath}
Considering $a(t)+b(t)h(x)$ as a new potential shape, the associated Schr\"odinger operator has eigenvalues equal to $v\,a(t)+ H_n(v\,b(t))$ as the monotonicty of $g$ ensures that $b(t)$ is positive.  The choice that $g''(h)>0$ implies that $g(h)$ lies above these tangential potentials: $f(x)=g(h(x))\geq a(t)+b(t)h(x)$ for all $x$.  Thus, by the Comparison Theorem \cite{thirring}, an almost immediate consequence of the variational min-max principle \cite{reed}, it follows that
\begin{displaymath}
E_n=F_n(v)\geq v\, a(t)+H_n(v\, b(t)).
\end{displaymath}
As this is valid for any point of contact, the family of tangential potentials forms a lower envelope for $g(h)$; maximizing over $t$ yields a lower bound for $F_n(v)$.
\medskip

In the complementary case, we assume $g$ is concave (i.e. $g''<0$), and a minimization over $t$ yields energy upper bounds.  The coupling parameter $v$ in the development so far is to be taken as a constant that could equally well have been absorbed into the definition of $f(x).$  The point of it at this stage is to show that with the same effort we approximate $F_n(v),$ not just $F_n(1).$  In the more succinct formulation of the method based on kinetic potentials, and outlined below in section 5, this coupling parameter in the target problem $\mathscr{H}$ will play a more essential r\^{o}le.
\medskip

One further relation that will be important is the following: since the potential $f(x)$ and its tangent $a(t)+b(t)h(x)$ touch at the point $x=t$, it is true for the case where $g$ is convex that $\min_x[f(x)-(a(t)+b(t)h(x))]=0$, leading to 
\begin{equation}\label{key}
\min_x[f(x)-b(t)h(x)]=a(t).
\end{equation}
A corresponding expression for $a(t)$ in the case where $g$ is concave requires a maximum over $x$. 
\medskip

To consider a specific example, we turn to estimating the eigenvalues of the quartic potential $f(x) = x^4$ by means of the harmonic oscillator $h(x)= x^2.$
\begin{example}
\normalfont Consider $\mathscr{H}=-\Delta+x^4$ and $\mathcal{H}=-\Delta+ux^2$. The transformation $g(h)=h^2$ satisfies the sufficient conditions of montonicity and convexity for a lower-bound estimate of $E_n$. We have that $a(t)=-t^4$ and $b(t)=2t^2$, and by an elementary scaling argument it can be shown that for the harmonic oscillator, $H_n(u)=u^{\half}\,H_n(1)=u^{\half}(2n+1)$. Thus we find that, for $n=0,1,\ldots$,
\begin{displaymath}
E_n\geq\max_t[a(t)+H_n(b(t))]=\max_t[-t^4+\sqrt{2t^2}(2n+1)]=\frac{3}{4}(2n+1)^{\frac{4}{3}}.
\end{displaymath} 
\end{example}

\section{The local energy theorem}
Introduced by Barta in \cite{barta} for vibrating membranes and then applied to quantum mechanics by Duffin in \cite{duffin} and Bartlett in \cite{bartlett}, the local energy theorem provides a method for estimating eigenvalues of a Schr\"{o}dinger operator $\mathscr{H}$ by looking at the local energy 
$$
\frac{\mathscr{H}\phi(x)}{\phi(x)}
$$
where $\phi$ is a trial function in the domain $\mathcal{D}(\mathscr{H})$.  Originally, the local energy theorem was formulated as technique for finding lower bounds to complement the well-known Rayleigh-Ritz upper bound.  However, the theorem can provide both upper and lower bounds:
\begin{theorem}
Let $\phi\in{C^2(\mathbb{R})}$ be such that $\phi$ undergoes no change of sign and vanishes only at isolated points.  If $v$ is large enough so that $vf(x)>E_0$ for sufficiently large $|x|$, then
\begin{displaymath}
\inf_{x}\left(\frac{\mathscr{H}\phi(x)}{\phi(x)}\right)\leq E_0 \leq \sup_{x}\left(\frac{\mathscr{H}\phi(x)}{\phi(x)}\right),
\end{displaymath}
where the points such that $\phi$ vanishes are excluded.
\end{theorem}
It is important to note that, in general, the local energy is not bounded.  One can only expect it to be either bounded above or bounded below.
\medskip

A rigorous proof of the lower-bound portion of this theorem first appears due to Barnsley in \cite{barnsley}.  Moreover, Barnsley extends the lower-bound result to the excited states in the following form:  
\begin{theorem}
Let $\phi\in{C^2(\mathbb{R})}$ be such that it undergoes exactly $n$ changes in sign and vanishes only at isolated points.  If $v$ is large enough so that $vf(x)>E_n$ for sufficiently large $|x|$, then, for $n=0,1,2,\ldots$,
\begin{displaymath}
\inf_{x\in{I}}\left(\frac{\mathscr{H}\phi(x)}{\phi(x)}\right) \leq E_n,
\end{displaymath}
where the points such that $\phi(x)$ vanishes are excluded.
\end{theorem}
\medskip

\noindent Subsequently, interest in the upper-bound estimate of the theorem re-emerged.  In particular, the works of Baumgartner \cite{baumgartner}, Schmutz \cite{schmutz}, and Thirring \cite{thirring} each analyze the theorem using different approaches.  However, to the knowledge of the present authors, an upper-bound local-energy estimate was never proven in the case of excited states.
\medskip

In more recent years, however, authors such as Mouchet \cite{mouchet1,mouchet2} and Handy \cite{handy} have further analyzed the groundstate local energy theorem and have identified two inherent deficiencies related to its practical use.  First, one has no {\it a priori} knowledge of whether a given trial function will yield an upper or a lower bound.  This will be addressed in the following section of the paper.  Second, there is no way systematically to tighten a bound once a trial function has been used.  Both Mouchet and Handy develop approaches that are suited for generating numerical estimates via the groundstate local energy theorem.
\medskip

Building on the previous example, we use the excited-state version of the local energy theorem to estimate the eigenvalues of the quartic potential $f(x) = x^4$ by means of the harmonic oscillator $h(x)= x^2.$
\begin{example}
\normalfont Consider $\mathscr{H}=-\Delta+x^4$ and $\mathcal{H}=-\Delta+tx^2$.  We may approximate $E_n$ by the wave function (involving the $n$th Hermite polynomial, $\mathfrak{H}_n(x)$) associated with $\mathcal{H}$, which is, indeed, of class $C^2$ in the $x$ variable for all $n$. Thus, using the local energy theorem with trial function $\phi_{n}(t;x)=\mathfrak{H}_n(t^{\frac{1}{4}}x)\exp({-\half\,{t^{\half}}\,x^2})$,
\begin{eqnarray*}
E_n&\geq&\max_t\left[\min_x\left[\frac{\mathscr{H}\phi_{n}(t;x)}{\phi_{n}(t;x)}\right]\right]\\&=&\max_t\left[\min_x\left[{t^{\half}}(2n+1)-tx^2+x^4\right]\right]\\&=&\frac{3}{4}(2n+1)^{\frac{4}{3}}.
\end{eqnarray*} 
\end{example}

\section{The coincidence}
We point out that in the previous example the bound obtained for the quartic potential's spectrum, $E_n\geq\frac{3}{4}(2n+1)^{\frac{4}{3}}$, by means of the local energy theorem is the same as the bound obtained in an earlier example by envelope theory.  This is by no means an isolated incident; in fact, we shall prove the following:
\begin{theorem}\label{coincidence}
Consider the Schr\"{o}dinger problems \refeq{Fn} and \refeq{base}.  Denote the bound obtained from the local energy theorem with trial function $\phi$ by $\overline{E}$ and the bound obtained from envelope theory with the potential $h$ by $\underline{E}$. Then $\overline{E}=\underline{E}$.
\end{theorem}
Note that the state $n$ is suppressed in the statement of this theorem as to indicate that it holds with either $g$ convex or concave when $n=0$, or with $g$ convex for the excited states.  Since the theorem holds for the case of $g$ convex for all $n=0,1,\ldots$, we present the proof with that set-up and reintroduce the notation that incorporates the state $n$.  
\medskip

Assuming that $g$ is convex, envelope theory gives us that
\begin{displaymath}
E_n\geq a(t)+ H_n(b(t))\Rightarrow E \geq\max_{t}[a(t)+ H_n(b(t))]\equiv \underline{E_n}
\end{displaymath}
and the local energy theorem gives us that 
\begin{displaymath}
E_n\geq\min_x\left[\frac{\mathscr{H}\phi_n(t;x)}{\phi_n(t;x)}\right]
\Rightarrow E_n\geq\max_{t}\left[\min_x\left[\frac{\mathscr{H}\phi_n(t;x)}{\phi_n(t;x)}\right]\right]\equiv\overline{E_n}. 
\end{displaymath}
Hence, using \refeq{key}, we have
\begin{eqnarray*}
\underline{E_n}&=&\max_{t}[a(t)+H_n(b(t))]
\\&=&\max_{t}[\min_x[f(x)-b(t)h(x)]+H_n(b(t))]
\\&=&\max_{t}[\min_x[f(x)-b(t)h(x)+H_n(b(t))]]
\\&=&\max_{t}\left[\min_x\left[f(x)-\frac{\phi''_n(t;x)}{\phi_n(t;x)}\right]\right]
\\&=&\max_{t}\left[\min_x\left[\frac{\mathscr{H}\phi_n(t;x)}{\phi_n(t;x)}\right]\right]
\\&=&\overline{E_n}.
\end{eqnarray*}
\qed
The idea of the proof is identical for the case of concave $g$, with the only change occurring from the switch between minima and maxima.
\medskip

This coincidence allows us to address an issue that was raised in the previous section: given a trial function $\phi\in\mathcal{D}$, we do not know {\it a priori} whether the local energy theorem will yield an upper or a lower bound.  By theorem 3, however, we {\it do} know whether the local energy theorem will yield an upper or a lower bound in the special case where $\phi$ is chosen to be the wavefunction to the problem \refeq{base} for some $h$ for which there exists a transformation such that $f=g(h)$ and $g$ satisfies the conditions \refeq{connection}.  Namely, theorem 3 says that an upper bound will be obtained if $g$ is concave and that a lower bound will be obtained if $g$ is convex. 
\medskip 

Moreover, as restricting the class of trial functions in the local energy theorem to those generated by Schr\"{o}dinger problems like \refeq{base} results in bounds equal to those obtained by envelope theory, the proof of theorem 3 implies that the complementary bound to theorem 2 holds for this restricted class of trial functions.  Namely, the following generalisation of the local energy theorem holds:
\begin{theorem}
Let $\phi_n$ be taken from the problem $\refeq{base}$ where $g$ is convex. Then, for $n=0,1,2,\ldots$,
\begin{displaymath}
E_n\leq\sup_{x\in{I}}\left(\frac{\mathscr{H}\phi_n(x)}{\phi_n(x)}\right),
\end{displaymath}
where the points such that $\phi_n(x)$ vanishes are excluded.
\end{theorem}

\section{Kinetic potentials and the the direct spectral formulation of the energy bound}
We express the envelope energy bounds entirely in terms of certain spectral functions so that the local--energy trial function $\phi(r)$ is no longer needed to obtain the spectral approximations available by means of the local-energy method.  We first define kinetic potentials as a representation for the discrete spectral data of the type of Schr\"{o}dinger operator $\mathscr{H} = -\Delta + v\,f(x)$ that concerns us.  This representation is specially designed to facilitate the analysis of the spectral relationship induced by a smooth transformation $f = g(h)$ of a base potential shape $h(x)$ whose associated discrete spectrum of $-\Delta + v\, h(x)$ is known.  We first define the concept of `kinetic potential' and then we show that these spectral objects admit a very simple general expression of the potential--envelope method. 

\subsection{Kinetic potentials}
Beginning with problem \refeq{Fn}, we can discuss the application of min-max by expressing the process in terms of `kinetic potentials' ({\it minimum mean iso-kinetic potential} \cite{hall3,hall4,hall5,hall6}) whereby the optimization is effected in two stages:  In the first stage, the kinetic energy is constrained to have the value $s = \langle -\Delta\rangle>0$ and yields the kinetic potential $\overline{f}_n(s)$ associated with the potential shape $f$ and the eigenvalue $n$; and, in the second stage the result $ [s + v\overline{f}_n(s)]$ is minimized over the kinetic energy $s >0.$  Thus: 
\begin{equation}\label{kp}
\overline{f}_n(s)=\inf_{\mathcal{D}_n}\sup_{\psi\in\mathcal{D}_n}{(\psi,f\psi)},\quad {\rm and} \quad E_n= F_n(v) =\min_{s>0}[s+v\overline{f_n}(s)],
\end{equation}
where $\mathcal{D}_n$ are $n$-dimensional subspaces of $\mathcal{D}$.  We started with the potential shape $f(r)$ in $\mathscr{H} = -\Delta + vf(r)$ and then the energy functions $\left\{F_n(v)\right\}$ of $\mathscr{H}$  are represented by the corresponding kinetic potentials $\left\{\overline{f}_n(s)\right\}$.  The following Legendre \cite{gelfand} transformation relations allow us to go back and forth $\{F_n\}\longleftrightarrow \{\overline{f}_n\} $ between these two sets of spectral functions:
\begin{equation}\label{kfun}
\begin{aligned}
s&=&F_n(v)-vF'_n(v),~~\overline{f}_n(s)=F'_n(v)\\
\frac{1}{v}&=&-\overline{f'}_n(s),~~\frac{1}{v}F_n(v)=\overline{f}_n(s) - s\overline{f'}_n(s).
\end{aligned}
\end{equation}
It is straightforward to prove by a variational argument \cite{hall5a} that $F_n''(v) < 0$, and we can show that the Legendre transformation \refeq{kp} implies $F_n''(v) \overline{f}''_n(s) = -\frac{1}{v^3}$; that is to say, $F_n(v)$ is concave and $\overline{f}_n(s)$ is convex.
\medskip

In the case of a pure powers with Hamiltonian $-\Delta + v\, \sgn(q) |x|^q$ and eigenvalues for $v = 1$ written in the form $E^{(q)}_{n}$ in $d= 1$ spatial dimension we have \cite{hall5}
\begin{displaymath}
f^{(q)}(x) = \sgn(q)\,|x|^q, ~~~ F^{(q)}_{n}(v) = E^{(q)}_n v^{\frac{2}{2 + q}}, ~~~ \overline{f}^{(q)}(s) = \left(\frac{P^{(q)}_n} {s^{\half}}\right)^{q},
\end{displaymath}
\noindent where $n= 0,1,2,\dots $ is the number of nodes in the eigenfunction and the $P^{(q)}_n$ numbers are given by:
\begin{displaymath}
P^{(q)}_n = \left(\frac{\left|E^{(q)}_n\right|}{1+q/2}\right)^{\frac{2+q}{2q}}\left(\frac{|q|}{2}\right)^{\half} ,   \quad q \neq 0,\, q > -2.
\end{displaymath}
For example, in dimension $d = 1$ we have for the harmonic oscillator $f(x) = x^2$, $E = 1+2n,$ and $P = n + \half$.  for the hydrogen atom in $d = 3$ dimensions we obtain $E^{(-1)}_{n\ell} = -\frac{1}{4(n+\ell)^2},$ and $P^{(-1)}_{n\ell} = (n+\ell).$  For these power potentials a further change of variables $r = P_{n}^{(q)}/s^{\half}$ recasts the relation \refeq{kp} between the energy and the potential into the following semi-classical form:
\begin{displaymath}
F_{n}^{(q)}(v) = \min_{r>0}\left[\left(\frac{P_n^{(q)}}{r}\right)^2 + \,v\,f(r)\right].
\end{displaymath}
\medskip

\subsection{Smooth transformations of potentials}
The potential envelope method has a very simple expression \cite{hall4,hall5,hall6} in terms of kinetic potentials the proof of which is essentially by an application of Jensen's inequality \cite{feller}:
\begin{theorem}
Consider the Schr\"{o}dinger operator $\mathscr{H} = -\Delta + v\, f$, where the attractive potential shape $f$ is a smooth transformation $f = g(h)$ of an attractive potential $h$, the coupling parameter $v > 0,$ and the transformation function $g(h)$ is increasing with definite convexity.  Then it follows that $\overline{f}_n(s) \approx g(\overline{h}_n(s))$, where if $g'' >0,$ $\approx $~ becomes $ \geq$, and if $g'' <0,$ $\approx$~ becomes $ \leq.$
\end{theorem}
\medskip

\noindent Hence the envelope approximation may be written
\begin{displaymath}
E_n = F_n(v) \approx \min_{x>0}\left[s  + v\, g\left(\overline{h}_n(s)\right)\right],
\end{displaymath}
where the approximation is a lower bound if $g$ is convex and an upper bound if $g$ is concave.
\medskip

\begin{example}
Consider $\mathscr{H}=-\Delta+v x^4$ and the envelope generator $\mathcal{H}=-\Delta+u x^2$.  As mentioned in example 1, we have that $H_n (u) = u^{\half}\,H_n(1) = u^{\half}\,(2n+1)$ so that the Legendre transformation relations \refeq{kfun} yield
\begin{displaymath}
s=\frac{u^{\half}\,(2n+1)}{2},\qquad\overline{h}_n(s) = \frac{(n+\half)^2}{s}.
 \end{displaymath}
Since the transformation function $g(h) = h^2$ is convex, the envelope approximation yields the lower bound:
 \begin{displaymath}
 E_n = F_n(v) \geq \min_{s  > 0}\left[s + v \left(\overline{h}_n (s)\right)^2 \right] = 
 \min_{s  > 0}\left[s + v\,\frac{(n+\half)^4}{s^2}\right] = \min_{r  > 0}\left[\left(\frac{n+\half}{r}\right)^2 + v\,r^4\right]=
 \frac{3}{4}\,v^{\frac{1}{3}}\,(2n+1)^{\frac{4}{3}}.
 \end{displaymath}
 \end{example}
The dependence of the energy bound on $v$ is the same as that derived by scaling arguments for the exact solution.

\section{Formulation in $d > 1$ dimensions}
\noindent The $d$-dimensional Schr\"odinger equation, in atomic units $\hbar=2m=1$, with
a spherically symmetric potential $V(r)$ can be written as
\begin{equation}\label{eq1}
\left[-\Delta_d +V(r)\right]\psi(r)=E\psi(r),
\end{equation}
where $\Delta_d$ is the $d$-dimensional Laplacian and $r^2=\sum_{i=1}^d x_i^2$. In order to transform (\ref{eq1}) to the $d$-dimensional spherical coordinates $(r, \theta_1,\theta_2,\dots,\theta_{d-1})$, we follow Sommerfeld \cite{sommerfeld} and Louck \cite{louck}, and separate variables using
\begin{displaymath}
\psi(\mb{r})=r^{(k-1)/2}u(r)Y_{l_{d-1}\dots l_1}(\theta_1\dots\theta_{d-1}),
\end{displaymath}
where $r = ||\mb{r}||$, $Y_{l_{d-1}\dots l_1}(\theta_1\dots\theta_{d-1})$ is a normalized spherical harmonic with characteristic value $l(l+d-2)$, $~\ell_{d-1} = l = 0,1,2,\dots$ is the angular-momentum quantum number, and $k=2l + d$. We then write  the radial Schr\"odinger equation in the form
\begin{equation}\label{eq3}
H u(r) = \left[-\frac{d^2}{dr^2}+ \frac{(k-1)(k-3)}{4r^2}+V(r)\right]u(r)=E u(r),\quad \int_0^{\infty} u^2(r)dr=1, \quad u(0)=0.
\end{equation}
We suppose that the potential $V(r)$ is less singular than the centrifugal term so that
$$u(r)\sim r^{\frac{1}{2}(k-1)},\quad\quad r\rightarrow 0.$$
We note that the Hamiltonian and boundary conditions of (\ref{eq3}) are invariant under the transformation 
$$(d,l)\rightarrow (d\mp2,l\pm 1).$$ Thus the energy remains unchanged if $k=2\ell+d$ and the number of nodes $n$ are given. We have \cite{hall5}
\begin{displaymath}
f^{(q)}(r) = \sgn(q)\,r^q, ~~~ F^{(q)}_{nk}(v) = E^{(q)}_{nk}\,v^{\frac{2}{2 + q}}, ~~~
 \overline{f}^{(q)}_{nk}(s) = \sgn(q)\left(\frac{P^{(q)}_{nk}} {s^{\half}}\right)^{q},
\end{displaymath}
\noindent where $n= 0,1,2,\dots $ is the number of nodes in the radial eigenfunction and the  $P^{(q)}_{nk}$ numbers are given by:
\begin{displaymath}
P^{(q)}_{nk} = \left(\frac{\left|E^{(q)}_{nk}\right|}{1+q/2}\right)^{\frac{2+q}{2q}}\left(\frac{|q|}{2}\right)^{\half} ,   \quad q \neq 0,\, q > -2.
\end{displaymath}
Thus we have in particular for the Coulomb potential $q= -1$ and the harmonic oscillator $q = 2$:
\begin{displaymath}
P^{(-1)}_{nk} = n+\ell+(d-1)/2  = n +(k-1)/2 \quad {\rm and}\quad P^{(2)}_{nk} = 2n+\ell + d/2 = 2n +k/2.
\end{displaymath}
\medskip

We now consider an explicit example, namely the linear combination of the Coulomb and the harmonic--oscillator potentials:
\begin{displaymath}
V(r)=v\, \left(-\frac{A}{r}+Br^2\right),
\end{displaymath}
where the positive coefficients $A$ and $B$ are both constant, and $v>0$ is a coupling parameter. $V(r)$ is at once a concave function $V(r) = g^{(1)}(r^2)$ of $r^2$ and a convex function $V(r) = g^{(2)}(-1/r)$ of $-1/r$.  Thus tangents to the $g$ functions are either shifted scaled oscillators above $V(r)$, or shifted scaled Coulomb potentials below $V(r)$.  We illustrate the situation graphically in Fig.~1 for the case $k = 2\ell + d = 7,$  $ n = 0.$  The resulting envelope energy-bound formulas are given by

\begin{figure}[ht]
\label{fig:fig1}
\centering
\includegraphics[width=7cm,height=5cm]{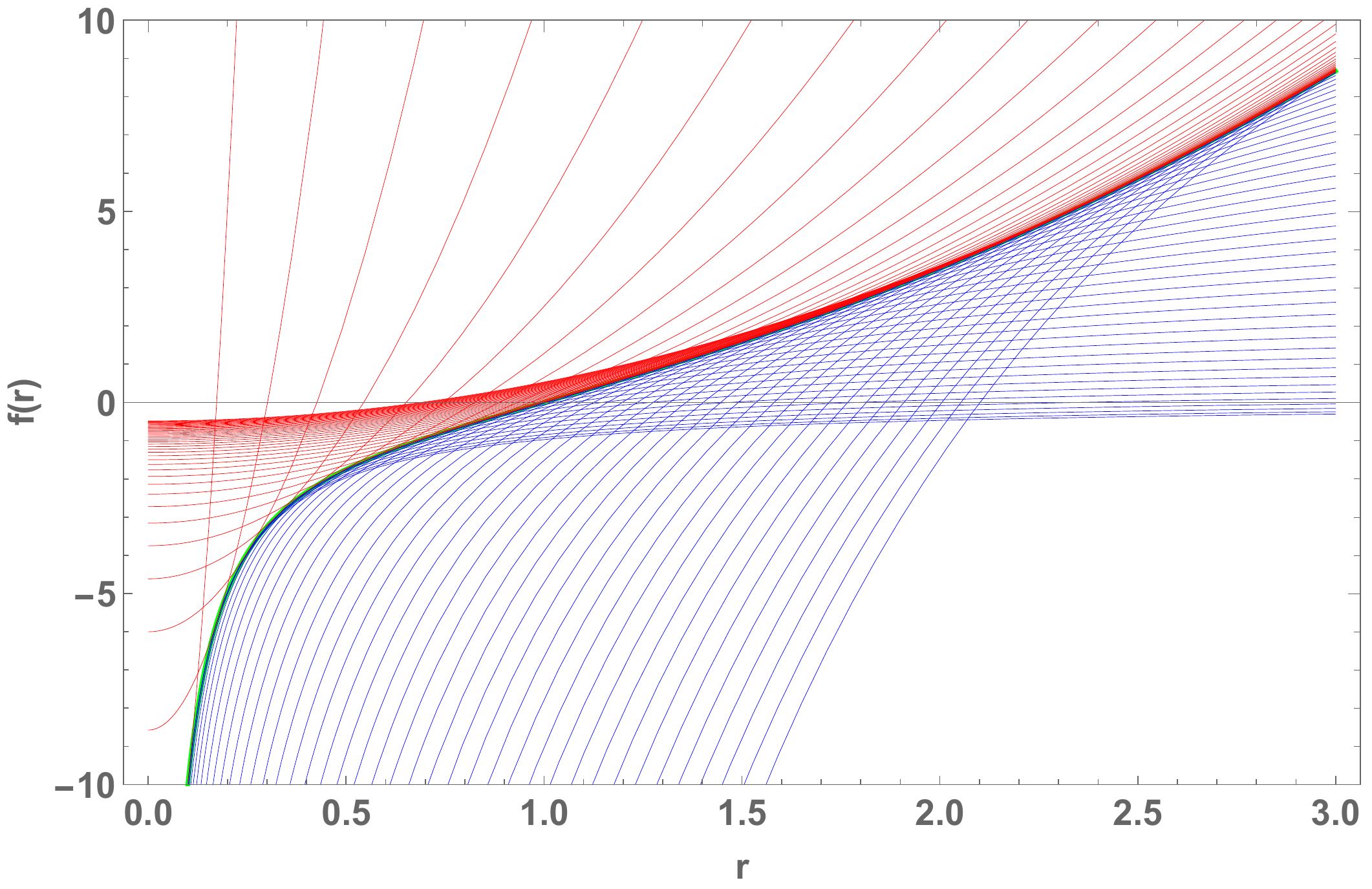} \qquad
\includegraphics[width=7.5cm,height=5.5cm]{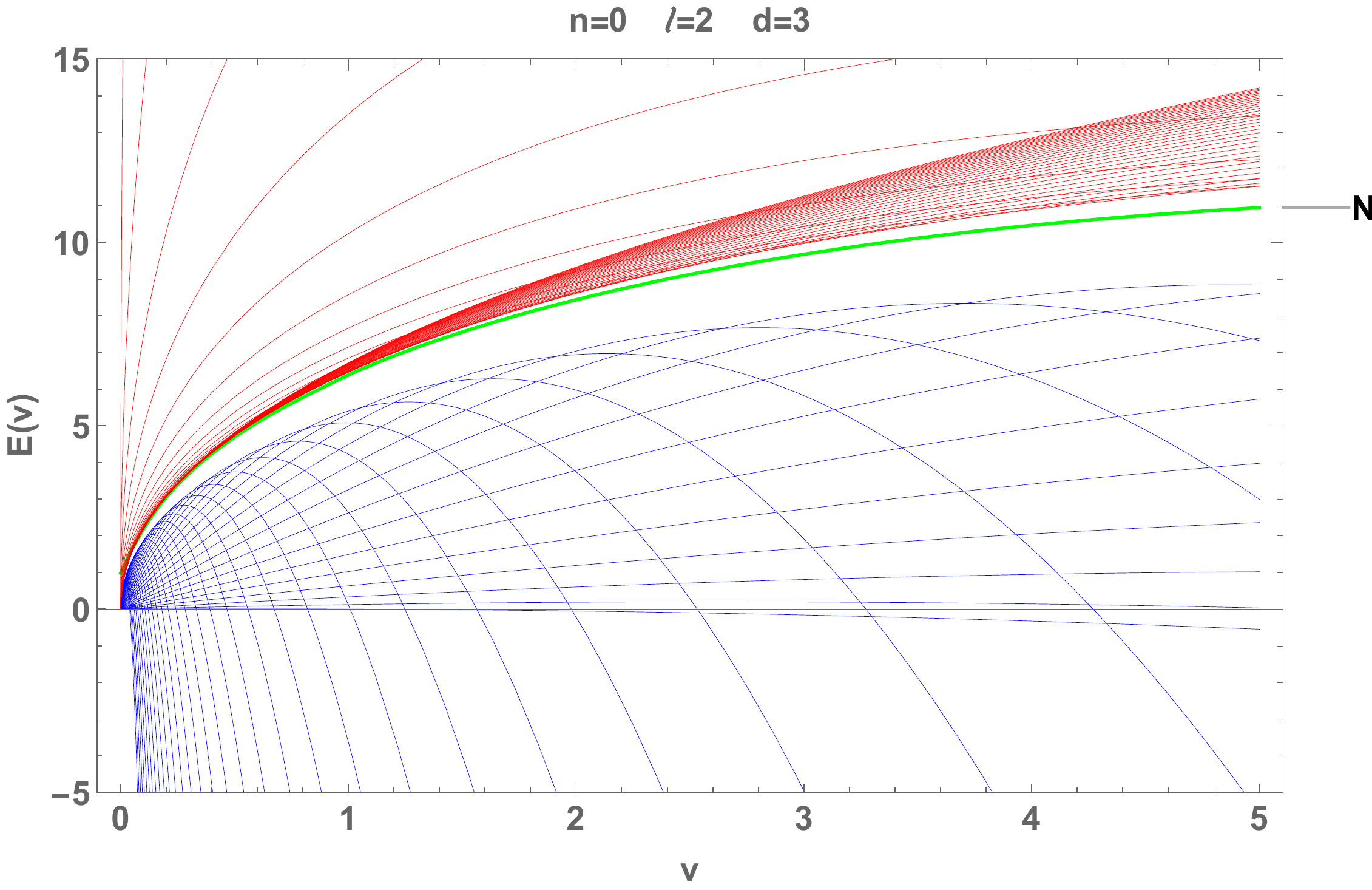}  
\caption{The potential $f(r) = -1/r + r^2$ is at once a convex transformation of the Coulomb potential $h_1(r) = -1/r$ and a concave transformation of the harmonic oscillator $h_2(r) = r^2.$  Thus $f(r)$ is the envelope curve of a lower Coulomb family and also an envelope curve of an upper oscillator family of potentials.  These families  generate corresponding families of spectral curves $-\Delta + v\,h_i(r) \longrightarrow F^{(i)}(v),$ where $v >0$ is the  coupling parameter.  Two such sets of spectral curves provide lower and upper bounds for the eigenvalues $F_{nk}(v)$ of the operator $-\Delta +v\,f(r).$  The eigenvalue considered here is $ n = 0$, $k = 2\ell + d = 7,$ with the potential parameters $ A = B = v =1$. An accurate spectral curve $F_{07}(v)$ was obtained by a numerical shooting method and is labelled `N'. }
\end{figure}
\begin{displaymath}
5.41553 < \min_{r > 0}\left[\left(\frac{P^{(-1)}_{nk}}{ r}\right)^2 \, + \, v\, \left(-\frac{A}{r} + B r^2\right) \right]\, \le \,E_{nk}\,\le\, \min_{r > 0}\,\left[\left(\frac{P^{(2)}_{nk}}{ r}\right)^2 \, + \, v\,\left( -\frac{A}{r} +B r^2\right)\right] < 6.46028,
\end{displaymath}
where the numerical bounds were obtained explicitly for the special case\nll  $\{A = B= v = 1, d = 3, \ell = 2, n = 0,\} \Rightarrow  k =  2\ell + d  =  7$.  It is clear that the lower energy bound has the Coulombic degeneracies, and the upper bound those of the harmonic oscillator.
\medskip

We now turn to the corresponding local-energy calculation of the bounds. Following section 3 above with the potential $V(r) = -1/r + r^2$ we observe that with $k=7,$ the power-of-$r$ factors in the wave functions are $r^{(k-1)/2} = r^3.$  Thus the lower envelope trial radial wave functions are of the form $\phi_1(t_1; r) = r^3 \exp(-t_1  r/2)$ and the upper oscillator trial wave functions are given by $\phi_2(t_2; r) = r^3 \exp(-t_2 r^2/2)$, where $t_1> 0$ and $t_2 > 0$ are parameters to be chosen optimally.  The Hamiltonian for the case we are considering is $H= -\left(\frac{\partial}{\partial r}\right)^2 + \frac{6}{r^2} -\frac{1}{r} + r^2$.  Thus the extrema of $w(t;r) = (H\phi(t;r))/ \phi(t;r)$ with respect to $r$ lead to the critical points given by $\frac{\partial}{\partial r} w(t;r) = 0$, which implies that the parameters $t_i$ can be taken to be functions of $r$.  We find $t_1(r) = (1 + 2 r^3)/3$ and $ t_2(r) = (1 + 1/(2 r^3))^{\half}$.  Thus the optimum energy estimates from the local energy theorem are obtained when we maximize the lower bound and minimize the upper bound to find
\begin{displaymath}
5.41553 < \max_{r>0}\left[w_1(t_1(r), r) +6/r^2 - 1/r + r^2\right] < E <  \min_{r>0}\left[w_2(t_2(r), r) +6/r^2 - 1/r + r^2\right] < 6.46028,
\end{displaymath}
in agreement with the bounds we obtained above by complementary sets of envelope curves exhibited in Fig.(1).
\vskip0.1true in
\section{Conclusion}
In this paper, we discuss an interesting relationship between two theories of spectral approximation for a Schr\"odinger operator $H = -\Delta + v\,f(r),$  which is assumed to be bounded below and to have some discrete eigenvalues for suitable choices of the coupling parameter $v >0.$  The potential shape $f(r)$ is written in the form $f(r) = g(h(r))$, where $g(h)$ is monotone increasing and of definite convexity $g''(h) \ne 0.$  Thus $f(r)$ is the envelope of a family of `tangential potentials' of the form $\{f^{(t)}(r) = a(t) + b(t)\,h(r)\}$, where $r = t$ is a point of contact between  the tangential potential shape $f^{(t)}(r)$ and the potential shape $f(r) $.  If $\phi(r)$ is an eigenfunction of the tangential operator $-\Delta + v\, f^{(t)}(r)$, then our analysis demonstrates that such a wave function, when used as a local-energy trial function, will yield a finite energy bound.  Moreover, if $E$ is an eigenvalue of $H$, then 
\begin{displaymath}
 \mybox{$g''>0$}  \Rightarrow E  \le \max_{r >0}\left[\frac{H\phi(r)}{\phi(r)}\right]~~~{\rm and}~~~~
 \mybox{$g''<0$}  \Rightarrow E \ge \min_{r >0}\left[\frac{H\phi(r)}{\phi(r)}\right].
\end{displaymath}
The trial wave function $\phi$ may be chosen to have the angular-momentum and nodal characteristics corresponding to the eigenvalue sought. We have featured the class of power-law potentials $f(r) = \sgn(q) r^q,\,q> -2,$ because these potentials provide some simple explicit illustrative examples. The kinetic-potential apparatus outlined in section~5A is convenient for the discussion but is not essential.

\medskip
 We hope that the existence of a geometrical approach (expressed in the present case by means of potential envelopes) yielding the  same results for spectral estimates of Schr\"odinger operators  as those given by the local energy theorem, may suggest  similar theoretical dualities for the large variety of other applications of the local-energy concept found in the literature.

 \medskip

 \begin{acknowledgments}
Partial financial support of his research under Grant No.~GP3438 from~the Natural
Sciences and Engineering Research Council of Canada is gratefully acknowledged.

 \end{acknowledgments}
 
 \vskip 2cm


\begin{thebibliography}{99} 



\bibitem{baumgartner}B. Baumgartner, {\it A class of lower bounds for Hamiltonian operators}, J. Phys. A {\bf 12}, 459 (1979).
\bibitem{barnsley}M. F. Barnsley, {\it Lower bounds for quantum mechanical energy levels}, J. Phys. A (London) {\bf 11}, 55 (1978).
\bibitem{barta}J. Barta, {\it Sur la vibration fondamentale d'une membrane}, C. R. Acad. Sci. Paris {\bf 204}, 472 (1937).
\bibitem{bartlett}J. H. Bartlett, {\it Helium wave equation}, Phys. Rev. {\bf 98}, 1067 (1955).
\bibitem{buis}F. Buisseret, C. Semay, and B. Silvestre--Brac, {\it Some equivalences between the auxiliary filed method and envelope theory}, J. Math. Phys. {\bf 50}, 032102 (2009).
\bibitem{duffin}R. J. Duffin, {\it Lower bounds for eigenvalues}, Phys. Rev. {\bf 71}, 827 (1947).
\bibitem{feller}W. Feller, {\it An introduction to probability theory and its applications, Volume II}, (Wiley and Sons, New York, 1971).
\bibitem{gelfand}I. M. Gelfand and S. V. Fomin, {\it Calculus of Variations}, (Dover Publications, New York, 1991).
\bibitem{hall1}R. L. Hall, {\it Energy trajectories for the N-boson problem by the method of potential envelopes}, Phys. Rev. D {\bf 22}, 2062 (1980).
\bibitem{hall3}R. L. Hall, {\it A geometrical theory of energy trajectories in quantum mechanics}, J. Math. Phys. {\bf 24}, 324 (1983).
\bibitem{hall4}R. L. Hall, {\it Kinetic potentials in quantum mechanics}, J. Math. Phys. {\bf 25}, 2078 (1984).
\bibitem{hall5}R. L. Hall, {\it Spectral geometry of power-law potentials in quantum mechanics}, Phys. Rev. A {\bf 39}, 5500 (1989).
\bibitem{hall6}R. L. Hall, {\it Envelope theory in spectral geometry}, J. Math. Phys. {\bf 34}, 2779 (1993).
\bibitem{hall7}R. L. Hall, {\it  Constructive inversion of energy trajectories is quantum mechanics}, J. Math. Phys. {\bf 40}, 699 (1999).
\bibitem{handy}C. R. Handy, {\it (Quasi)-convexification of Barta's (multi-extrema) bounding theorem}, J. Phys. A {\bf 39}, 3425 (2006).
\bibitem{louck} J. D. Louck,  J. Mol. Spectrosc. 4 (1960) 298-333.
\bibitem{mouchet1}A. Mouchet, {\it A differential method for bounding the ground state energy}, J. Phys. A {\bf 38}, 1039 (2005).
\bibitem{mouchet2}A. Mouchet, {\it Upper and lower bounds for an eigenvalue associated with a positive eigenvector}, J. Math. Phys. {\bf 47}, 022109 (2006).
\bibitem{reed}M. Reed and B. Simon, {\it Methods of Modern Mathematical Physics IV: Analysis of Operators}, (Academic Press, San Diego, 1978).
\bibitem{schmutz}M. Schmutz, {\it The factorization method and ground state energy bounds}, Phys. Lett. A {\bf 108}, 195 (1985).
\bibitem{sommerfeld} A. Sommerfeld {\it Partial Differential Equations in Physics}, (Academic Press, New York, 1949).  [The Laplacian in $d$ dimensions is discussed in Appendix IV, p227.]
\bibitem{thirring}W. Thirring, {\it A Course in Mathematical Physics 3: Quantum Mechanics of Atoms and Molecules}, (Springer, New York/Wien, 1990).

\end{thebibliography}
\end{document}